\documentclass[12pt,draftclsnofoot,peerreview,onecolumn]{IEEEtran}
\usepackage{psfrag}
\usepackage{amssymb}
\usepackage{amsmath}
\usepackage{cite}
\usepackage{graphics}
\usepackage{graphicx}
\usepackage{epsfig}
\usepackage{subfigure}
\usepackage{amscd}
\usepackage{color, soul}
	\title{Overview of Full-Dimension MIMO in LTE-Advanced Pro}
	
	\author{
		Hyoungju Ji, Younsun Kim, and Juho Lee, Samsung Electronics, Korea \\
		Eko Onggosanusi, Younghan Nam, and Jianzhong Zhang, Samsung Research America \\
		Byungju Lee, Purdue University\\
		Byonghyo Shim, Seoul National University \\
		\vspace{0.8cm}
		\thanks{To appear in IEEE Communications Magazine}
	}
\begin{document}	
\maketitle

\begin{abstract}
Multiple-input multiple-output (MIMO) systems with a large number of basestation antennas, often called massive MIMO, have received much attention in academia and industry as a means to improve the spectral efficiency, energy efficiency, and processing complexity of next generation cellular system.
Mobile communication industry has initiated a feasibility study of massive MIMO systems to meet the increasing demand of future wireless systems. Field trials of the proof-of-concept systems have demonstrated the potential gain of the Full-Dimension MIMO (FD-MIMO), an official name for %{massive} 
the MIMO enhancement in 3rd generation partnership project (3GPP). 3GPP initiated standardization activity for the seamless integration of this technology into current 4G LTE systems. %A study item, a process done before a formal standardization process, has been completed in June 2015, and the follow-up {work item process will be finalized shortly} for the formal standardization of Release 13.
In this article, we provide an overview of the FD-MIMO system, with emphasis on the discussion and debate conducted on the standardization process of Release 13. We present key features for FD-MIMO systems, a summary of the major issues for the standardization and practical system design, and performance evaluations for typical FD-MIMO scenarios.
\end{abstract}

%%%%%%%%%%%%%%%%%%%%%%%%%%%%%%%%%%%%%%%%%%%%%%%%%%%%%%%%%%%%%%%%%%%%%%%%%%%%%%%
%%%%%%%%%%%%%%%%%%%%%%%%%%%%%%%%%%%%%%%%%%%%%%%%%%%%%%%%%%%%%%%%%%%%%% ABSTRACT
	
	%\begin{keywords}
	% Do we need it ?
	%\end{keywords}

	% \begin{center} \bfseries EDICS Category: ??? \end{center}
	% \fi
	
	\IEEEpeerreviewmaketitle
	\setcounter{page}{2}

	%%%%%%%%%%%%%%%%%%%%%%%%%%%%%%%%%%%%%%%%%%%%%%%%%%%%%%%%%%%%%%%%%%%%%%%%%%%%
	\section{Introduction}
	%%%%%%%%%%%%%%%%%%%%%%%%%%%%%%%%%%%%%%%%%%%%%%%%%%%%%%%%%%%%%%%%%%%%%%%%%%%%%
	\label{sec:Introduction}
Multiple-input multiple-output (MIMO) systems with a large number of basestation antennas, often referred to as \emph{massive MIMO systems}, have received much attention in academia and industry as a means to improve the spectral efficiency, energy efficiency, and processing complexity \cite{Massive}. %{The wisdom behind the massive MIMO systems is that when the number of basestation antennas goes to infinity, multiuser interference caused by the downlink user co-scheduling and uplink multiple access approaches to zero, resulting in a dramatic increase in the throughput with relative simple transmitter and receiver operations.}
While the massive MIMO technology is a promising technology, there are many practical challenges and technical hurdles down the road to the successful commercialization. These include design of low-cost and low-power basestation with acceptable antenna space, improvement in the {fronthaul} capacity between radio and control units, acquisition of high dimensional channel state information (CSI), and many others. Recently, 3rd generation partnership project (3GPP) standard body initiated the standardization activity to employ tens of antennas at basestation with an aim to satisfy the spectral efficiency requirement of future cellular systems \cite{FDMIMO1, FDMIMO2}.	
Considering the implementation cost and complexity, and also the timeline to the real deployment, 3GPP decided to use tens of antennas with a two dimensional (2D) array structure as a starting point. Full-Dimension MIMO (FD-MIMO), the official name for %{massive} 
the MIMO enhancement in 3GPP, targets the system utilizing up to 64 antenna {ports} at the transmitter side.
Recently, field trials of the proof-of-concept FD-MIMO systems have been conducted successfully\cite{test1}. A study item, a process done before a formal standardization process, has been completed in June 2015, and the follow-up  {work item process will be finalized soon} for the formal standardization of Release 13 (Rel. 13).\footnote{LTE-Advanced Pro is the LTE marker that is used for the specifications from Release 13 onwards by 3GPP.}
		
{The purpose of this article is to provide an overview of the FD-MIMO systems with an emphasis on the discussion and debate conducted on the standardization process of Rel. 13. We note that preliminary studies addressed the feasibility of 2D array antenna structure and performance evaluation in ideal pilot transmission and feedback scenarios \cite{FDMIMO1, FDMIMO2}. This work is distinct from these in the sense that we put our emphasis on describing realistic issues in the standardization process, including TXRU architectures, beamformed CSI-RS, 3D beamforming, details of CSI feedback, and performance evaluation in realistic FD-MIMO scenarios with new feedback schemes. 
	
	%Specifically, we describe key features for FD-MIMO systems, which are not covered in} \cite{FDMIMO1}, {summarize the main {results} for the standardization process and system design, provide performance evaluation results for the {new CSI feedback schemes}, and also make some comments on the future direction.}

\section{Key Features of FD-MIMO Systems}
In this section, we discuss {key features of FD-MIMO systems. These include a large number of basestation antennas, 2D active antenna array, 3D channel propagation, and new pilot transmission with CSI feedback. } {In what follows, we will use LTE terminology exclusively: enhanced node-B (eNB) for basestation, user equipment (UE) for the mobile terminal, and reference signal (RS) for pilot signal.}
\subsection{{Increase the number of transmit antennas}}
	\label{sec:Features}
One of the main features of FD-MIMO systems distinct from the MIMO systems of the current LTE and LTE-Advanced standards is to use a large number of antennas at eNB.
In theory, as the number of eNB antennas $N_T$ increases, cross-correlation of two random channel realizations goes to zero \cite{Massive} so that the inter-user interference in the downlink can be controlled via a simple linear precoder. %\st{and multiuser interference in the uplink can be eliminated via a simple receive combiner.} 
Such benefit, however, can be realized only when the perfect CSI is available at the eNB. While the CSI acquisition in time division duplex (TDD) systems is relatively simple due to the channel reciprocity, such is not the case for frequency division duplex (FDD) systems. Note that in the FDD systems, time variation and frequency response of the channel are measured via the downlink {RSs} and then sent back to the eNB after the quantization.
Even in TDD mode, one cannot solely rely on the channel reciprocity because the measurement at the transmitter does not capture the downlink interference from neighboring cells or co-scheduled UEs. As such, downlink {RSs} are still required to capture the channel quality indicator (CQI) for the TDD mode, and thus the downlink {RS} and the uplink CSI feedback are essential for both duplex {modes}. Identifying the potential issues of CSI acquisition and developing the proper solutions are, therefore, of great importance for the successful commercialization of FD-MIMO systems. Before we go into detail, we briefly summarize two major problems related to the CSI acquisition. % process are as follows.

\begin{itemize}
	\item \textbf{Degradation of CSI accuracy}: One well-known problem for the MIMO systems, in particular for FDD-based systems, is that the quality of CSI is affected by the limitation of feedback resources. As the CSI distortion increases, quality of the multiuser MIMO (MU-MIMO) precoder to control the inter-user interference is degraded and so will be the performance of the FD-MIMO systems. In general, the amount of CSI feedback, determining the quality of CSI, needs to be scaled with $N_T$ to control the quantization error so that the overhead of CSI feedback increases in FD-MIMO systems. % would be a concern. \st{ in massive MIMO regime.}
		
	\item \textbf{Increase of pilot overhead}: An important problem related to the {CSI acquisition} %{large-scale antennas} 
	at eNB, yet to be discussed separately, is the pilot overhead problem. UE performs the channel estimation using the RS transmitted from the eNB. Since RSs need to be assigned in an orthogonal fashion, RS overhead typically grows linearly with $N_T$. For example, if $N_T=64$, RS will occupy approximately $48\%$ of resources, eating out % will be used for RS\st{ in LTE systems}. %\footnote{Resource block (RB) is a minimum scheduling unit for data transmission and it comprises 168 resource elements (REs) = 14 symbols $\times$ 12 subcarriers. When the first 3 symbols are used for control channels, 132 REs are available for data symbols and RSs.} <--- remove for word lengh
     substantial amount of {downlink} resources for the data transmission.
\end{itemize}

\subsection{{2D active antenna system (AAS)}}

	{Another} interesting feature of the FD-MIMO system is an introduction of the active antenna with 2D planar array. In the active antenna-based systems, gain and phase are controlled by the active components, such as power amplifier (PA) and low noise amplifier (LNA), attached to each antenna element. In the 2D structured antenna array, one can control the radio wave on both vertical (elevation) and horizontal (azimuth) direction so that the control of the transmit beam in 3D space is possible.
This type of wave control mechanism is also referred to as the \emph{3D beamforming}. Another important benefit of 2D {AAS} is that it can accommodate a large number of antennas without increasing the deployment space. For example, when 64 linear antenna arrays are deployed in a horizontal direction, under the common assumption that the antenna spacing is half wavelength ($\frac{\lambda}{2}$) and the system is using LTE carrier frequency (2 GHz), it requires a horizontal room of $3$m. Due to the limited space on a rooftop or mast, this space would be burdensome for most of the cell sites. In contrast, when antennas are arranged in a square array, relatively small space is required for 2D antenna array (e.g., 1.0 $\times$ 0.5m  with dual-polarized 8 $\times$ 8 antenna array).

\subsection{{3D channel environment}}

When basic features of the FD-MIMO systems are determined, the next step is to design a system maximizing performance in terms of throughput, spectral efficiency, and peak data rate {in the realistic channel environment}. There are various issues to consider in the design of {practical} systems, such as investigation and characterization of the realistic channel model for the performance evaluation. While the conventional MIMO systems consider the propagation in the horizontal direction only, FD-MIMO systems employing 2D planar array should consider the propagation in both vertical and horizontal direction. To do so, geometric structure of the transmitter antenna array and propagation effect of the 3D positions between the eNB and UE should be reflected in the channel model. % Also, effect of the vertical propagation caused by the height difference between the transmitter and receiver should be considered in the channel {propagation and the CSI} feedback design. 
Main features of 3D channel propagation obtained from real measurement are as follows \cite{3D}:

%, design of new transmitter architecture for supporting 2D active antenna array, and pilot transmission and feedback strategy in accordance with these changes.
%
%
%In particular, following factors need to be highlighted on top of the conventional 2D channel propagation model: % in the design of FD-MIMO systems. <-- word length
\begin{itemize}
	%\item The channel model should consider the effect of angular spread in the vertical direction.	
	\item {Height and distance-dependent line-of-sight (LOS) channel condition: LOS probability between eNB and UE increases with the UE's height and also increases when the distance between eNB and UE decreases. }
	\item {Height-dependent pathloss: UE experiences less pathloss on a  higher floor (e.g., 0.6dB/m gain for macro cell and 0.3dB/m gain for micro cell).}
	\item {Height and distance-dependent elevation spread of departure angles (ESD): When the location of eNB is higher than the UE, ESD decreases with the height of the UE. It is also observed that the ESD decreases sharply as the UE moves away from the eNB.}
	%
    %
    %

    %  
	%\item New transmitter architectures, often referred to as transceiver unit (TXRU) architecture, need to be added in the design of the transmitter. By TXRU architecture, we mean a hardware connection between the baseband signal path and antenna array elements. In the active antenna system, patch antennas and active devices are integrated on the printed circuit board (PCB) so that one can easily design paths between TXRUs and antenna elements. Since this architecture facilitates the control of phase and gain in both digital and analog domain, more accurate control of the beamforming direction is possible. One thing to note is that the conventional codebook cannot measure the CSI for the beamformed transmission so that new RS transmission and channel feedback mechanism supporting the beamformed transmission should be introduced (See Section \ref{sec:Issues} for details).
\end{itemize}

		\begin{figure*}
			\centering
			\includegraphics[width=165mm]{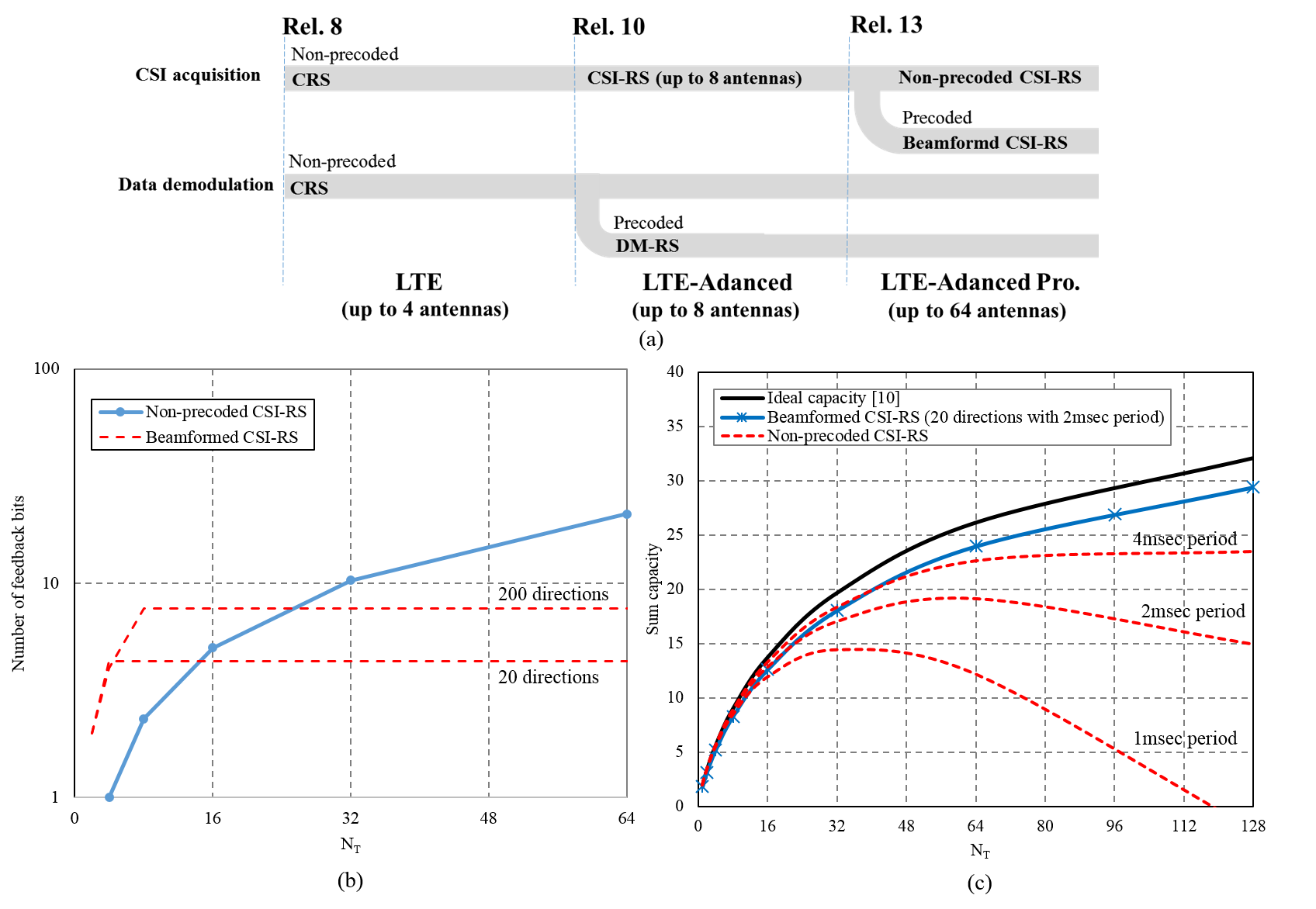}
			\caption{MIMO evaluation: (a) RS evolution in LTE systems, (b) uplink feedback overhead (SNR=$10$dB \cite{Jindal}), (c) MU-MIMO capacity with considering CSI-RS overhead (ideal CSI and ZFBF MU-precoding with 10 UEs and SNR=$10$dB).}
			\label{fig:new}
		\end{figure*}
		
\subsection{{RS transmission for CSI acquisition}}

 	{From the LTE to LTE-Advanced, there has been substantial improvement in the RS scheme for MIMO systems (see Fig. 1(a)). From the common RS (CRS) to the channel state information RS (CSI-RS), various RSs  to perform the CSI acquisition have been introduced. While these are common to all users in a cell and thus un-precoded, the demodulation RS (DM-RS) is UE-specific (i.e., dedicated to each UE) so that it is precoded by the same weight applied for the data transmission. Since the DM-RS is present only on time/frequency resources where the UE is scheduled, this cannot be used for CSI measurements} \cite{MUMIMO1}.

    One of the new features of the FD-MIMO systems is to use a beamformed RS, called beamformed CSI-RS, for the CSI acquisition. Beamformed RS transmission is a channel training technique that uses multiple precoding weights in spatial domain. In this scheme, UE picks the best weight among transmitted and then feeds back its index. This scheme provides many benefits over non-precoded CSI-RS, in particular when {$N_T$ is large}. Some of the benefits are summarized as follows:
\begin{itemize}
	\item \textbf{Less uplink feedback overhead:} In order to maintain a rate comparable to the case with perfect CSI, feedback bits used for the channel vector quantization should be proportional to $N_T$ \cite{Jindal}. Whereas, the amount of feedback for the beamformed CSI-RS scales logarithmic with the number of RSs $N_B$ %($\log_2 N_B$, where $N_B$ is the number of beamformed CSI-RS)
	since this scheme only feeds back an index of the best beamformed CSI-RS. Thus, as depicted in Fig. \ref{fig:new}(b), the benefit of beamformed CSI-RS is pronounced when $N_T$ is large.
	%	this scheme is required for beamformed CSI-RS with directing beams in $N_B$ dominant pathes. 
	%For example, with 20 beamformed CSI-RS, uplink feedback overhead is smaller than than non-preocded one when more than 14 transmit antennas are used (see Fig. \ref{fig:new}(b)). %{Since the number of dominant paths is invariant with the number of transmission antennas, additional overhead would be minimal.}	
	\item \textbf{Less downlink pilot overhead: } %Since non-precoded CSI-RS is common for all UEs, pilot for each antenna should be assigned in an orthogonal fashion. {%For non-precoded CSI-RS, the transmit antennas occupy orthogonal resources which are common for UEs in a cell. 
	When the non-precoded CSI-RS is used, pilot overhead increases with $N_T$, resulting in a substantial loss of the sum capacity in the FD-MIMO regime (see Fig. \ref{fig:new}(c)).
	%As shown in Fig. \ref{fig:new}(c), When the number of transmit antennas increases, pilot overhead eats out the sum capacity with non-precoded CSI-RS. % and thus capacity is rather decreased} (see Figure \ref{fig:new}(c)). 
	Whereas, pilot overhead of the beamformed CSI-RS is proportional to $N_B$ and independent of $N_T$ so that the rate loss of the beamformed CSI-RS is marginal even when $N_T$ increases.
	% irrespective of the number of antennas. Due to this reason, rate efficiency of the beamformed CSI-RS increases with $N_T$. 
		%when $N_B>N_T$, capacity with beamformed CSI-RS outperforms non-precoded CSI-RS}\cite{Goldsmith}.
	\item \textbf{Higher quality in RS:} If the transmit power is $P$ watt, $P/N_T$ watt is needed for each non-precoded CSI-RS transmission, while $P/N_B$ watt is used for the beamformed CSI-RS. For example, when $N_T=32$ and $N_B=12$, {beamformed CSI-RS provides 4.3dB gain in signal power over the non-precoded CSI-RS.\footnote{In 3D channel model, the typical number of multi-paths (clusters) is 12 \cite{3D}.}}% at most 6dB SNR difference can be occurred and this lead more accurate CSI acquisition and small mean square error with beamformed CSI-RS.}
\end{itemize}

{%For CSI-RS transmission, various beamforming strategies can be used in digital and analog domain with different types of AAS. 
In order to support the beamformed CSI-RS scheme, new transmitter architecture called transceiver unit (TXRU) architecture has been {introduced}. By TXRU architecture, we mean a hardware connection between the baseband signal path and antenna array elements. %{In the active antenna system, patch antennas and active devices are integrated on the printed circuit board (PCB) so that one can easily design paths between TXRUs and antenna elements.} 
Since this architecture facilitates the control of phase and gain in both digital and analog domain, more accurate control of the beamforming direction is possible. One thing to note is that the conventional codebook cannot measure the CSI of the beamformed transmission so that a new channel feedback mechanism supporting the beamformed transmission is required {(see Section III.D for details)}.

	%%%%%%%%%%%%%%%%%%%%%%%%%%%%%%%%%%%%%%%%%%%%%%%%%%%%%%%%%%%%%%%%%%%%%%%%%%%%
	\section{System Design and Standardization of FD-MIMO Systems}

%	\subsection{Status of MIMO Standardization}
	\label{sec:status}

 The main purpose of the {Rel. 13 study item} is to identify key issues to support up to 64 transmit antennas placed in the form of a 2D antenna array.
 %Very recently, this study item has been completed and a detailed discussion for the standardization will be initiated shortly.
 %
 Standardization of the systems supporting up to 16 antennas is an initial target of Rel. 13 and issues to support more than 16 antennas will be discussed in subsequent releases. In {the study item} phase, there has been extensive discussion to support 2D array antennas, elaborated {TXRUs}, enhanced channel measurement and feedback schemes, and also an increased number of co-scheduled users (up to eight users). Among these, an item tightly coupled to the standardization is the CSI {measurement and} feedback mechanism. In this subsection, we discuss the deployment scenarios, {antenna configurations}, TXRU structure, new RS strategy, and feedback mechanisms.
\label{sec:Issues}
%\subsubsection{Overview}
%\label{sec:overview}

%
%

	\subsection{Deployment scenarios}	
%	\subsubsection{Deployment scenarios}
	\label{sec:scenarios}

	For the {design and evaluation} of FD-MIMO systems, a realistic scenario in which antenna array and UEs are located in different height is considered. To this end, two typical deployment scenarios, viz., 3D urban macro scenario (3D-UMa) and 3D urban micro (3D-UMi), are {introduced} (see Fig. \ref{fig:scenarios}). In the former case, transmit antennas are placed over the rooftop, and in the latter case, they are located below the rooftop. In case of 3D-UMa, diffraction over the rooftop is a dominant factor for the propagation so that down-tilted transmission in the vertical direction is desirable (see Fig. \ref{fig:scenarios}(b)). In fact, by transmitting beams with different steering angles, eNB can separate channels corresponding to multiple UEs. In the 3D-UMi scenario, on the other hand, the location of users is higher than the height of the antenna so that direct signal path is dominant (see Fig. \ref{fig:scenarios}(c)). In this scenario, both up and down-tilting can be used to schedule UEs in different floors. Since the cell radius of the 3D-UMi scenario is typically smaller than that of 3D-UMa, LOS channel condition is predominant, and thus more UEs can be co-scheduled without increasing the inter-user interference \cite{3D}. Although not as strong as the 3D-UMi scenario, LOS probability in the 3D-UMa scenario also increases when the distance between eNB and UE decreases.
		%Whereas, in 3D-UMa scenario, LOS probability increases as the distance between eNB and UE decreases, however, this is less dominant than 3D-UMi scenario.% inter-user interference can be reduced when multi-user (MU) precoding is applied. From this reason, more UEs can be co-scheduled. %Since inter-user interference is reduced ing each others. \st{in the vertical direction.}
   
	\subsection{Antenna configurations}
	\label{sec:antenna}   
	{Unlike the conventional MIMO systems relying on the passive antenna, systems based on the active antenna can {dynamically} control the gain of an antenna element by applying the weight of low-power amplifiers attached to each antenna element. Since the radiation pattern depends on the antenna arrangement, such as the number of the antenna elements and antenna spacing, the antenna system should be modeled in an element-level. % and its configuration parameters should be selected properly.}
	%\begin{itemize}
	%	\item The antenna system should be modeled in an element-level. Unlike the conventional MIMO systems relying on the passive antenna, systems based on the active antenna can dynamically control the gain of an antenna element by applying the weight to low-power amplifiers attached to each antenna element. Since the radiation pattern depends on the antenna arrangement such as the number of the antenna elements and antenna spacing, new precoding strategy to support the element-level antenna structure is required.
	%\end{itemize}
	%Beam pattern in 2D array antenna structure should be designed in a way to maximize the cell coverage and beamforming gain. 
	As shown in Fig. \ref{fig:antennas}(a), there are three key parameters characterizing the antenna array structure $(M, N, P)$: the number of elements $M$ in vertical direction, the number of elements $N$ in horizontal direction, and the polarization degree $P$ ($P=1$ is for co-polarization and $P=2$ is for dual-polarization). {As a benchmark setting}, 2D planar array using dual polarized antenna ($P=2$) configuration with $M=8$ ($0.8\lambda$ spacing in vertical direction) and $N=4$ ($0.5\lambda$ spacing in horizontal direction) is {suggested}.\footnote{Note that the total number of antenna elements in this setup is the same as that of 8Tx antennas in conventional systems {and thus FD-MIMO eNB can provide backward compatibility} \cite{FDMIMOTR}. {The vertical configuration is to ensure the same cell coverage and the horizontal configuration is for the conventional MIMO operation for LTE.}}
    %Since the angular spread in vertical domain is smaller than in horizontal, large aperture size can provide sharper beam with less scattering in vertical domain . %
   %  angular spread in decreases with the height of UE (reflecting diffraction angles from above-rooftop propagation). It is also observed that the ESD decreases sharply as a UE moves away from the eNB.
    In this setting, null direction, an angle to make the magnitude of beam pattern to zero, for the elevation beam pattern is 11$^{\circ}$ and that for the horizontal beam pattern is 30$^{\circ}$ (see Fig. \ref{fig:antennas}(c)).
    Since the null direction in the vertical domain is much smaller than that of the horizontal domain, scheduling UEs in the vertical domain is more effective in controlling the inter-user interference. % one can control the interuser interference effectively by scheduling UEs in the vertical domain.
    Also, a tall or fat array structure ($M\gg N$ or $M\ll N$) is favorable since it will generate a sharp beam but it might be less flexible in the situation where the surrounding environment is changed. % (e.g., user location, cell radius, building height, and antenna height) is changed. % provide sharper beam (less inter-beam interference) than square structure with the same number of antenna elements, but those will regress beamforming controllability from 3D to 2D.
    Further, large antenna spacing is not always a desirable option since it can increase the inter-cell interference due to the narrow beamforming for cell edge UEs (this phenomenon is called \emph{flash-light effect}). For this reason, in a real deployment scenario, the design parameters should be carefully chosen by considering various factors, such as user location, cell radius, building height, and antenna height.
   			\begin{figure*}
   				\centering
   				\includegraphics[width=165mm]{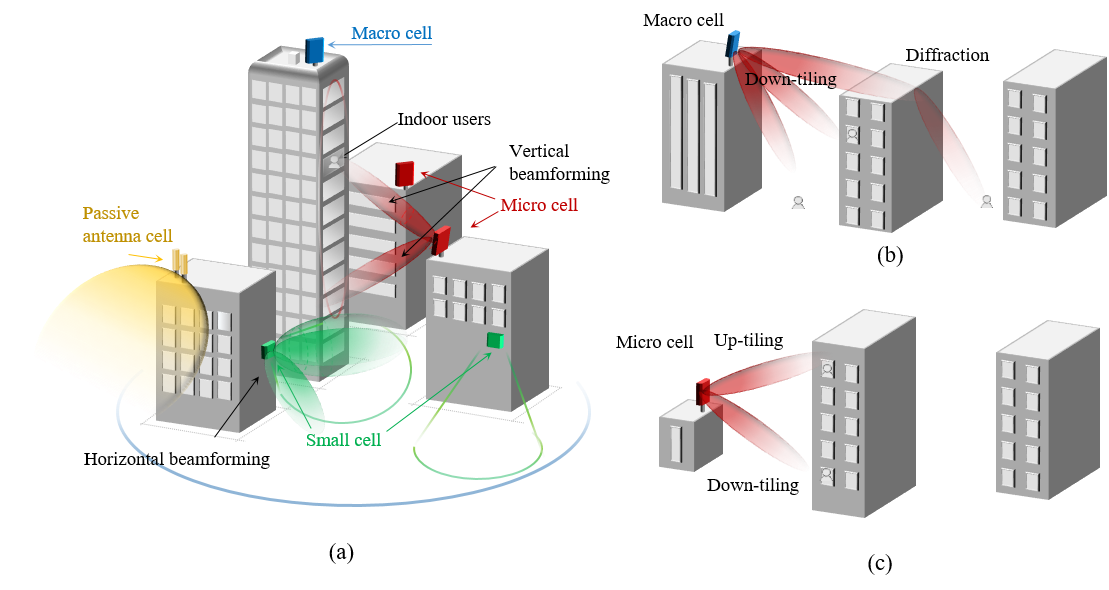}
   				\caption{FD-MIMO deployment scenarios: (a) 3D macro cell site (placed over the rooftop) and 3D micro cell site (placed below the rooftop) with small cell, (b) beamforming for 3D macro cell, and (c) beamforming in 3D micro cell.}
   				\label{fig:scenarios}
   			\end{figure*}
	
	\subsection{TXRU architectures}
	\label{sec:Architectures}
	%%%%%%%%%%%%%%%%%%%%%%%%%%%%%%%%%%%%%%%%%%%%%%%%%%%%%
	%%%%%%%%%%%%%%%%%%%%%%%%%%%%%%%%%%%%%%%%%%%%%%%%%%%%%
	\begin{figure*}
		\centering
		\includegraphics[height=190mm]{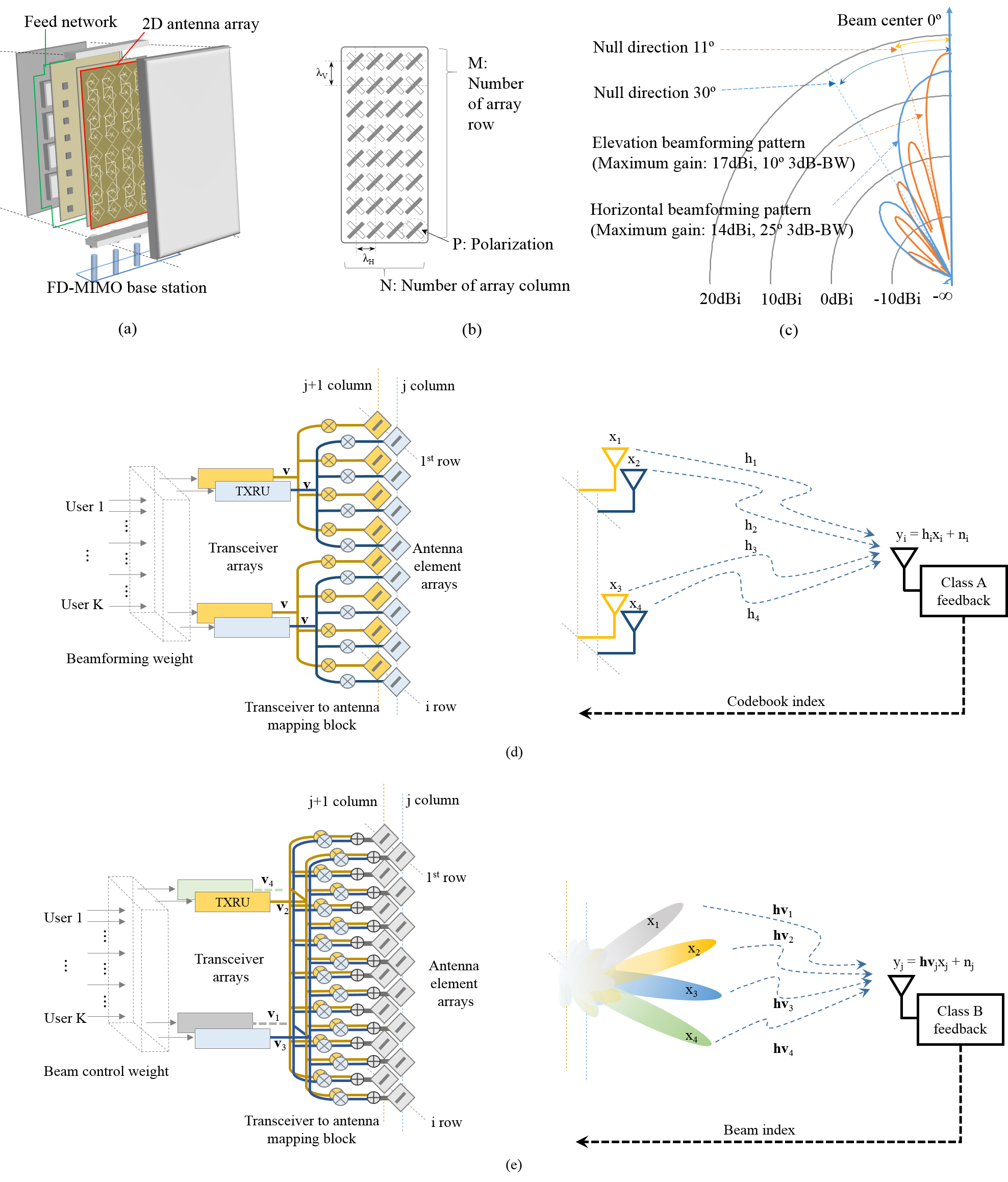}
		\caption{FD-MIMO systems: (a) concept of FD-MIMO systems, (b) 2D array antenna configuration, (c) vertical and horizontal beamforming patterns, (d) array partitioning architecture with the conventional CSI-RS transmission, and (e) array connected architecture with beamformed CSI-RS transmission.}
		\label{fig:antennas}
	\end{figure*}
	%%%%%%%%%%%%%%%%%%%%%%%%%%%%%%%%%%%%%%%%%%%%%%%%%%%%

   As mentioned, one interesting feature of the active antenna systems is that each TXRU contains PA and LNA so that eNB can control the gain and phase of an individual antenna element.
   In order to support this, a power feeding network between TXRUs and antenna elements called \emph{TXRU architecture} is introduced \cite{rank4AAS}. %can be implemented .
    TXRU architecture consists of three components: TXRU array, antenna array, and radio distribution networks (RDN). A role of the RDN is to deliver the transmit signal from PA to antenna array elements and the received signal from antenna array to LNA. Depending on the CSI-RS transmission and feedback strategy, two representative options, \emph{array partitioning} and \emph{array connected architecture}, are suggested. The former is for the conventional codebook scheme and the latter is for the beamforming scheme. %schemes-RS transmission strategies, i.e., extension of conventional non-precoded CSI-RS and beamformed CSI-RS, are suggested.    
    
    %
    %
    %
  %  	\begin{figure*}
  %  		\centering
  %  		\includegraphics[width=155mm]{fig3.png}
  %  		\caption{FD-MIMO transceiver architectures and \st{pilot} (CSI-RS) transmission strategy: (a) array partitioning architecture with the conventional CSI-RS transmission and (b) array connected architecture with beamformed CSI-RS transmission. One-to-one mapping between TXRU to antenna port is assumed.}
  %  		\label{fig:architecture}
  %  	\end{figure*}	
    In the array partitioning architecture, antenna elements are divided into multiple groups and each TXRU is connected to one of them (see Fig.  \ref{fig:antennas}(d)). Whereas, in the array connected structure, RDN is designed such that RF signals of multiple TXRUs are delivered to the single antenna element. To mix RF signals from multiple TXRUs, additional RF combining circuitry is needed as shown in Fig. \ref{fig:antennas}(e).
    The difference between the two can be better understood when we discuss the transmission of the CSI-RS. In the array partitioning architecture, $N_T$ antenna elements are partitioned into $L$ groups of TXRU and orthogonal CSI-RS is assigned for each group. 
    Each TXRU transmits its own CSI-RS so that the UE measures the channel $h$ from the CSI-RS observation $y= {h}x+n$. %Specifically, 
    %
    % remove for word count reduction 
    %
    %\footnote{When identical weight vector $\mathbf{v}=[{v}_1,\ldots,{v}_{L}]^T$ is applied to construct radiation pattern for each TXRU, vertical antenna ports are introduced together with horizontal ports. Effectively, 4$\times 0.8\lambda$ = 3.2$\lambda$ spaced vertical antenna ports can be configured as in Fig. \ref{fig:architecture}(a). Effective vertical antenna spacing can be reduced by placing more TXRUs in each column.} For each group, UE measures the channel vector $\mathbf{h}$ from the CSI-RS observation $\mathbf{y}= \mathbf{h}x+\mathbf{n}$.
    %
    %
In the array connected architecture, each antenna element is connected to $L^{'}$ (out of $L$) TXRUs and orthogonal CSI-RS is assigned for each TXRU. 
% with ${N_{T} \frac{L^{'}}{L}}=N_c$ dimension weight vector (see Fig. \ref{fig:antennas}(e) for $L^{'} = L$ and $N_c=N_T$). 
Denoting $\mathbf{h} \in \mathbb{C}^{1 \times N_c}$ as the channel vector and $\mathbf{v} \in \mathbb{C}^{N_c \times 1}$ as the precoding weight (${N_{T} \frac{L^{'}}{L}}=N_c$) for each beamformed CSI-RS, the beamformed CSI-RS observation is $y= \mathbf{hv}x+n$ and the UE measures the precoded channel $\mathbf{hv}$ from this. 
%Thus, UE measures the precoded channel $\mathbf{hv}$ from the beamformed CSI-RS observation $y= \mathbf{hv}x+n$ where $\mathbf{h} \in \mathbb{C}^{1 \times N_c}$ is channel vector and $\mathbf{v} \in \mathbb{C}^{N_c \times 1}$ is precoding weight for CSI-RS transmission.
%
%
Due to the narrow and directional CSI-RS beam transmission with a linear array, SNR of the precoded channel is maximized at the target direction.\footnote{SNR $=\frac{|\mathbf{hv}(\phi)|^2}{\sigma^2}$, where $\phi$ is the beam direction and $\sigma^2$ is the noise power.} % and can be expressed with a few dominant taps which effectively experiences less scattering to UEs. 
%In the next subsection, we discuss more on the CSI-RS transmission strategy.  <-- word count reduction

	\subsection{New CSI-RS transmission strategy}
	\label{sec:Pilots}	
 
    In the standardization process, two CSI-RS transmission strategies, i.e., extension of the conventional non-precoded CSI-RS and the beamformed CSI-RS, are suggested.    
    In the first strategy, UE observes the non-precoded CSI-RS transmitted from {each of partitioned antenna arrays} (see Fig.\ref{fig:antennas}(d)). By sending the precoder maximizing the properly designed performance criterion to the eNB, UE can adapt to the channel variation.
    In the second strategy, eNB transmits multiple beamformed CSI-RS (we call it \emph{beam} for simplicity) {using  connected arrays architecture.} Among these, UE selects the preferred beam and then feeds back its index. When the eNB receives the beam index, the weight corresponding to the selected beam is used for the data transmission. %In doing so, eNB obtains the channel direction information.

   Overall {downlink} precoder for data transmission $\mathbf{W}_{\textrm{data}} $ and CSI-RS transmission $\mathbf{W}_{\textrm{rs}} $  can be expressed as
   %%%%%%%%%%%%%%%%%%%%%
	\begin{equation}\label{pilots_general}
		\mathbf{W}_{\textrm{data}} = \mathbf{W}_{T}\mathbf{W}_{\textrm{rs}} \ \ \text{and} \ \ \mathbf{W}_{\textrm{rs}} = \mathbf{W}_P \mathbf{W}_{U},
	\end{equation}
   %%%%%%%%%%%%%%%%%%%%%%
	where $\mathbf{W}_{T} \in \mathbb{C}^{N_T \times L}$ is the precoder between TXRU and the antenna element, $\mathbf{W}_P \in \mathbb{C}^{L \times N_P}$ is the precoder between the CSI-RS port and the TXRU ($N_P$ is the number of antenna ports), and $\mathbf{W}_{U} \in \mathbb{C}^{N_{P}\times r}$ is the precoder between data channel to CSI-RS port. %Note that  CSI-RS signal is $\mathbf{x}_p = \mathbf{W}_T \mathbf{W}_{P} \mathbf{s}_p$ and the data signal is $\mathbf{x}_d =  \mathbf{W}_{T} \mathbf{W}_P \mathbf{W}_{U} \mathbf{s}_d$ where $\mathbf{s}_p$ is $L$ pilot symbols and $\mathbf{s}_d$ is rank $r$ data symbols. %Note that the weight applied to the CSI-RS is the product of $\mathbf{W}_{T}$ and $\mathbf{W}_P$ while the weight applied to the data transmission is the product of $\mathbf{W}_{T}$, $\mathbf{W}_P$, and. 

	In the following, we summarize details of two strategies.

	\begin{itemize}
		\item Conventional CSI-RS transmission: One option to maximize the capacity is to do one-to-one mapping of the TXRU and the CSI-RS resource (i.e., $\mathbf{W}_P =\mathbf{I}_{N_{TXRU}}$). To achieve the same coverage for each CSI-RS resource, an identical weight $\mathbf{v}$ is applied to $L$ groups.\footnote{In this paper, we assume that discrete Fourier transform (DFT) weights are used as $\mathbf{W}_{T}$ for mapping between TXRU and antenna elements for simplicity. For example, $\mathbf{W}_{T}$ can be expressed as $\mathbf{W}_{T} = [\mathbf{v} \ \mathbf{v}; \mathbf{v} \ \mathbf{v}]$ in Fig. \ref{fig:antennas}(d).} %This operation requires both CSI-RS resources as well as the number of TXRUs. When tens of TXRUs are used, CSI-RS overhead becomes a serious concern.
		Each UE measures the CSI-RS resources and then chooses the preferred codebook index $i^*$ maximizing the channel gain for each subband:
%
%%%%%%%%%%%%%%%%%%%%%%%%%%
		\begin{equation}
		i^* = \arg \max_{i} \| \bar{\mathbf{h}}^{H} \mathbf{W}_{U}^{i} \|_2^2,
		\end{equation}
%%%%%%%%%%%%%%%%%%%%%%%%%%
%
where $\|\mathbf{a}\|_2=\sqrt{\sum\limits_{i} |{a}_i|^2}$ and $\bar{\mathbf{h}}={\mathbf{h}}/{\|\mathbf{h}\|_2}$ is the estimated channel direction vector, and $\mathbf{W}_{U}^{i}$ is the $i$th precoder between the data channel and CSI-RS ports. {This scheme is called class-A CSI feedback.}
%
		%Since this approach increases the number of antenna ports in both horizontal and vertical axis, some modification to reduce the feedback overhead is needed.
%
%
%
		\item Beamformed CSI-RS transmission: In order to acquire the spatial angle between the eNB and UE, eNB transmits multiple beamformed CSI-RSs. Let $N_B$ be the number of CSI-RSs, then we have $\mathbf{W}_{T} = [\mathbf{v}_1 \mathbf{v}_2 \dots \mathbf{v}_{N_{B}}]$ where $\mathbf{v}_i \in \mathbb{C}^{N_T \times 1}$ is the 3D beamforming weight for the $i$th beam. %For example, one can use a \emph{grid-of-beam} approach for $\mathbf{W}_{T}$.
		For example, when the rank-1 beamforming is applied, we have $\mathbf{W}_{P}={\mathbf{1}_{N_{B}}}$ and $\mathbf{W}_{U}=1$.
		Among all possible beams $\mathbf{v}_1, ... , \mathbf{v}_{N_B}$, UE selects and feeds back the best beam index $j^*$ maximizing the received power:
		\begin{equation}
		j^{\ast} = \arg \underset{j}{\max} |\bar{\mathbf{h}}^H \mathbf{v}_j |^2.
		\end{equation}
        {This scheme is called class-B CSI feedback.} %While an index of the codebook maximizing the performance criterion is sent to eNB in the {class A} approach, an index of the best beam is fed back to eNB in this approach.
        %
        % <-- modify for word count reduction
        % While the conventional feedback mechanism picks up the codebook maximizing the performance criterion and then sends the index of it to eNB, this scheme feeds back an index of the best beam to eNB.
        Under the rich scattering environment, dominant paths between eNB and UE depend on the direction and width of the transmit signal. In the multiple-input single-output (MISO) channel, for example, the channel vector in an angular domain is expressed as $\mathbf{h}=\sum_{i} e_r\mathbf{e}_t(\phi_i)^*$, where $e_r=1$ and $\mathbf{e}_t(\phi_i)=[1 \ \ e^{-j2\pi\gamma\phi_i } \ ... \ \ e^{-j2\pi(N_T-1)\gamma\phi_i}]^T$ is the spatial signature of the transmitter ($\phi_i$ is direction of $i$th path and $\gamma$ is normalized antenna spacing) \cite{Tae}.
        % in the $i$th path direction of $\phi$ with normalized antenna spacing $\gamma$ for receiver and transmitter, respectively  
        When the RS is transmitted in a direction $\phi_j$, the beamforming weight would be $\mathbf{v}=\mathbf{e}_t(\phi_j)$ so that the resulting beamformed channel is readily expressed as one or at most a few dominant taps ($\mathbf{e}_t(\phi_i)^T\mathbf{e}_t(\phi_i)\approx 0$ when $i\neq j$). In fact, by controlling the weight applied to CSI-RS, the effective dimension of the channel vector can be reduced so that the feedback overhead can be reduced substantially. % an efficient feedback scheme. 
	\end{itemize}
	%In addition,  when eNB changes beamforming weight, UE does not recognize whether channel is fluctuated by fading or weight changes. %  semi-static or dynamically and thus channel in beamformed CSI-RS vary with both fading and weight change. 
	%In view of this, the primary goal of {CSI feedback with beamformed CSI-RS transmission} is to design a system such that UE can easily track the weight change. Clearly, this is in contrast to the role of conventional MIMO feedback adapting to the channel variation. To support this operation, the eNB needs to report the weight changes periodically or UE needs to identify the weight change by itself. 
	In Table \ref{tab:comparison}, we summarize two CSI-RS transmission schemes discussed in the FD-MIMO.
	
	\begin{table*}[t]\footnotesize
		\caption{Comparison between CSI-RS transmission and CSI feedback classes}
		\label{tab:comparison}
		\begin{tabular}{p{3.5cm}||p{6cm}|p{6cm} }
			\hline
			Category      &  Class-A CSI feedback (Conventional CSI-RS) &  Class-B CSI feedback (Beamformed CSI-RS) \\ \hline \hline
			Feedback design & Need to design codebook for 2D antenna layout and feedback mechanism for adapting channel variation & Need to devise a method to feed back beam index for adapting both weight changes and channel variation \\ \hline
			UL Feedback overhead & Depend on resolution of codebook and the number of antennas & Depend on the number of operating beam $N_B$\\ \hline
			CSI-RS overhead & Require $N_T$ CSI-RS resources  & Scale linearly with the number of beam $N_B$\\ \hline
			Backward compatibility & Supportable with virtualization between & Supportable with vertical 1D beamforming \\ 
			 & TXRUs and antenna ports & weight\\ \hline
			Forward compatibility & Scalable to larger TXRU system if CSI-RS resources are allowed & Scalable to larger TXRU system if long-term channel statistics are acquired\\
			\hline
		\end{tabular}
	\end{table*}	

	\subsection{CSI feedback mechanisms for FD-MIMO systems}
	\label{sec:feedback}
	%%%%%%%%%%%%%%%%%%%%%%%%%%%%%%%%%%%%%%%%%%%%%%%%%%%%%%%%%%%%%%%%%%%%%%%%%%%%
	 In the study item phase, various RS transmission and feedback schemes have been proposed. {As shown in Fig.} \ref{fig:new}{, capacity and overhead of class-A and class-B feedback schemes are more or less similar in the initial target range ($N_t =16$) so that Rel. 13 has decided to support both classes. %Through the heavy debate, both classes are supposed to be acceptable for Release 13 standard. 
	 In this subsection, we briefly describe the CSI feedback schemes associated with TXRU architectures.} {Among various schemes, composite codebook and beam index feedback have received much attention as main ingredients for class-A and class-B CSI feedback. The rest will be considered in a future release.}

	\textbf{Composite codebook}: In this scheme, overall codebook is divided into two (vertical and horizontal codebooks) and thus the channel information is separately delivered to the eNB. By combining two codebooks (e.g., Kronecker product of two codebooks $\mathbf{W}_{U}$ = $\mathbf{W}_{U,V} \otimes \mathbf{W}_{U,H}$), eNB reconstructs whole channel information. Considering that the angular spread of the vertical direction is smaller than that of the horizontal direction, one can reduce the feedback overhead by setting a relatively long reporting period to the vertical codebook. %The conventional LTE codebook can be reused for horizontal codebook, but it might be better to newly design the vertical codebook to achieve better tradeoff between performance and feedback overhead.	

	\textbf{Beam index feedback}: To obtain the UE's channel direction information (CDI) from beamformed CSI-RSs, eNB needs to transmit multiple beamformed CSI-RSs. When the channel rank is one, feedback of a beam index and corresponding CQI is enough.
	Whereas, when the channel rank is two with dual-polarized antennas, co-phase information is additionally required for adapting channel orthogonalization between layers. For example, once eNB obtains the CDI, this can be used for the beamforming vector of two-port CSI-RS and each CSI-RS port is mapped to the different polarized antennas. UE then estimates and feeds back short-term co-phase information between two ports. % and CQI without PMI reports.
	%A drawback of this scheme is that a large number of beams are needed to obtain an accurate CDI.
	 %One way to reduce pilot overhead is to transmit multiple spatially separated beams on the same CSI-RS resource. Alternative way is to employ aperiodic and trigger-based CSI-RS instead of the conventional (periodic) CSI-RS. % that used in the conventional systems.

	\textbf{Other CSI feedback schemes}: In the \emph{partial CSI-RS transmission}, CSI-RS overhead can be reduced by partitioning the 2D antenna array into horizontal and vertical ports, say $N_{H}$ ports in the row and $N_{V}$ ports in the column. In doing so, the total number of CSI-RS can be reduced from $N_H \times N_V$ to $N_H + N_V$. Overall channel information can be reconstructed by exploiting spatial and temporal correlation among antenna elements \cite{Lee}.
	In the \emph{adaptive CSI feedback} scheme, benefits of the beamformed and non-precoded CSI-RS transmission can be combined. First, in order to acquire long-term channel information, eNB transmits $N_T$ non-precoded CSI-RSs. After receiving sufficient long-term channel statistics from UE, eNB determines spatial direction roughly and then transmits the beamformed CSI-RSs used for short-term and subband feedbacks. The \emph{flexible codebook} scheme can support various 2D antenna layouts without increasing the number of codebooks. In this approach, one master codebook is designed for a large number of TXRUs, say 16 TXRUs, and the specific codebook (e.g., ($2\times8$), ($4\times4$), or ($1\times16$)) is derived based on this. To support this, the eNB needs to send the layout information via separate signaling. % and UE reconstructs the actual codebook using this information.

	\section{Performance of FD-MIMO System}
	\label{sec:performance}
	In order to observe the potential gain of the FD-MIMO systems, we perform system-level simulations under the realistic multicell environment. In our simulations, we test two typical deployment scenarios (3D-UMa and 3D-UMi) with 2-tier hexagonal layout. As a performance metric, we use spectral efficiency for cell average and cell edge. Detailed simulation parameters are provided in Table \ref{tab:simulation}.
    We first investigate the system performance of FD-MIMO systems with two types of antenna configurations. For type I and II configurations, $(M, N, P) = (8, 4, 2)$ and $(M, N, P) = (32, 4, 2)$ are used, respectively. In the type II configuration,  antenna spacing is set to four times larger than the spacing of type I. To investigate the effect of antenna structure, the ideal feedback under the full buffer traffic model (each user has an unlimited amount of data to transmit) is used. % In case of the full buffer model, each user has unlimited amount of data to transmit. 
    In Fig. \ref{fig:performance}(a), we plot the throughput of the conventional LTE systems with 8Tx ($N_V {\mkern-2mu\times\mkern-2mu} N_H = 1{\mkern-2mu\times\mkern-2mu}8$) and FD-MIMO systems with 16, 32, and 64Tx ($N_V{\mkern-2mu\times\mkern-2mu}N_H = 2{\mkern-2mu\times\mkern-2mu}8, \ 4{\mkern-2mu\times\mkern-2mu}8,$ and $\ 8{\mkern-2mu\times\mkern-2mu}8$), where $N_V$ and $N_H$ are the number of CSI-RS in vertical and horizontal dimensions, respectively. %As type II configuration, we plot the throughput with same ($N_V {\mkern-2mu\times\mkern-2mu} N_H$) antenna configuration under $(M, N, P) = (8N_V, 4, 2)$.
    %FD-MIMO systems achieve substantial gain over the conventional LTE MIMO systems, resulting in $62$\% and $130$\% gain for cell average and $105$\% and $484$\% gain for cell edge of type I and II, respectively.  
    This result shows that both antenna configurations provide a large gain over the conventional 8Tx in LTE-A, resulting in $105$\% (type I) and $484$\% (type II) gain at cell edge, respectively. Due to the sufficient antenna spacing, cross-correlation between channels becomes negligible, and thus the spectral efficiency of type II increases linearly with the TXRU, resulting in $30$\% (cell average) and $70$\% gain (cell edge) when the number of TXRUs is doubled \cite{Rapa}. However, due to the insufficient antenna spacing, the spectral efficiency of type I configuration does not scale linearly with the number of TXRUs. %\footnote{Noting that cross-correlation between channel decreases exponentially with the antenna spacing increase \cite{Rapa}}.

	We {next investigate} the system performance under the finite traffic model (e.g., FTP model) where  each UE with distinct arrival time receives a file with finite size. %User packet throughput is used as metric %In each subband, eNB selects a pair of UEs maximizing sum-capacity.
	As a performance metric, we use a user packet throughput, the number of successively received packets during the transmission period. In order to support the backward compatibility and also perform fair comparison among schemes under test, we employ the conventional MMSE-based channel estimation. % from both DM-RS and CSI-RS.  %use 50\%-tile of user packet throughput (representing cell average performance) and 5\%-tile user packet throughput (representing cell edge performance).
	% are considered: 50\%-tile user throughput represents cell average performance and 5\%-tile user throughput represents cell edge performance.
	In our simulations, the {following} CSI feedback strategies are considered. %  for the type I configuration.
	\begin{itemize}
		\item \textbf{Conventional 8Tx LTE Systems}: Rel. 10 LTE-A feedback mechanism using 8TX codebook is used. The implicit feedback (RI, horizontal and vertical PMIs, CQI) is used for the CSI feedback. %(RI, PMI, CQI) is used for the CSI feedback.
		\item \textbf{FD-MIMO systems with:}
		\begin{itemize}
			\item \textbf{Non-precoded CSI-RS}: A composite codebook of horizontal and vertical codebooks is used. In case of 16Tx with ($N_V{\mkern-2mu\times\mkern-2mu}N_H$=2$\times$8) antenna configuration, the codebook is generated via the Kronecker product of 2Tx and 8Tx LTE codebooks. The implicit feedback is used for the CSI feedback.
			\item \textbf{Beamformed CSI-RS scheme I}: Beam index feedback is used. Four beams are used to represent the vertical angles  ($N_B=4$). Each UE reports the best beam index (BI) and corresponding CQI.
			\item \textbf{Beamformed CSI-RS scheme II}: The eNB transmits both non-precoded and beamformed CSI-RS. UE feeds back long-term CSI (RI, long-term PMI) using the non-precoded CSI-RSs and reports the short-term CSI (BI, CQI) using the beamformed CSI-RSs ($N_B=4$). The precoding weight of beamformed CSI-RS is changed based on the long-term PMI. %for short-term CSI reporting (RI-L, PMI-L, PMI-S, CQI).
			%\item FD-MIMO systems with hybrid CSI-RS: The eNB configures both non-precoded CSI-RS and beamformed CSI-RS. Long-term CSI is fed back from non-precoded CSI-RS. The eNB changes precoding weight of beamformed CSI-RS for short-term CSI reporting (RI-L, PMI-L, PMI-S, CQI).			
		\end{itemize}

	\end{itemize}	
	%%%%%%%%%%%%%%%%%%%%%%%%%%%%%%%%%%%%%%%%%%%%%%%%%%%%%
	\begin{figure*}
		\centering
		\includegraphics[width=165mm]{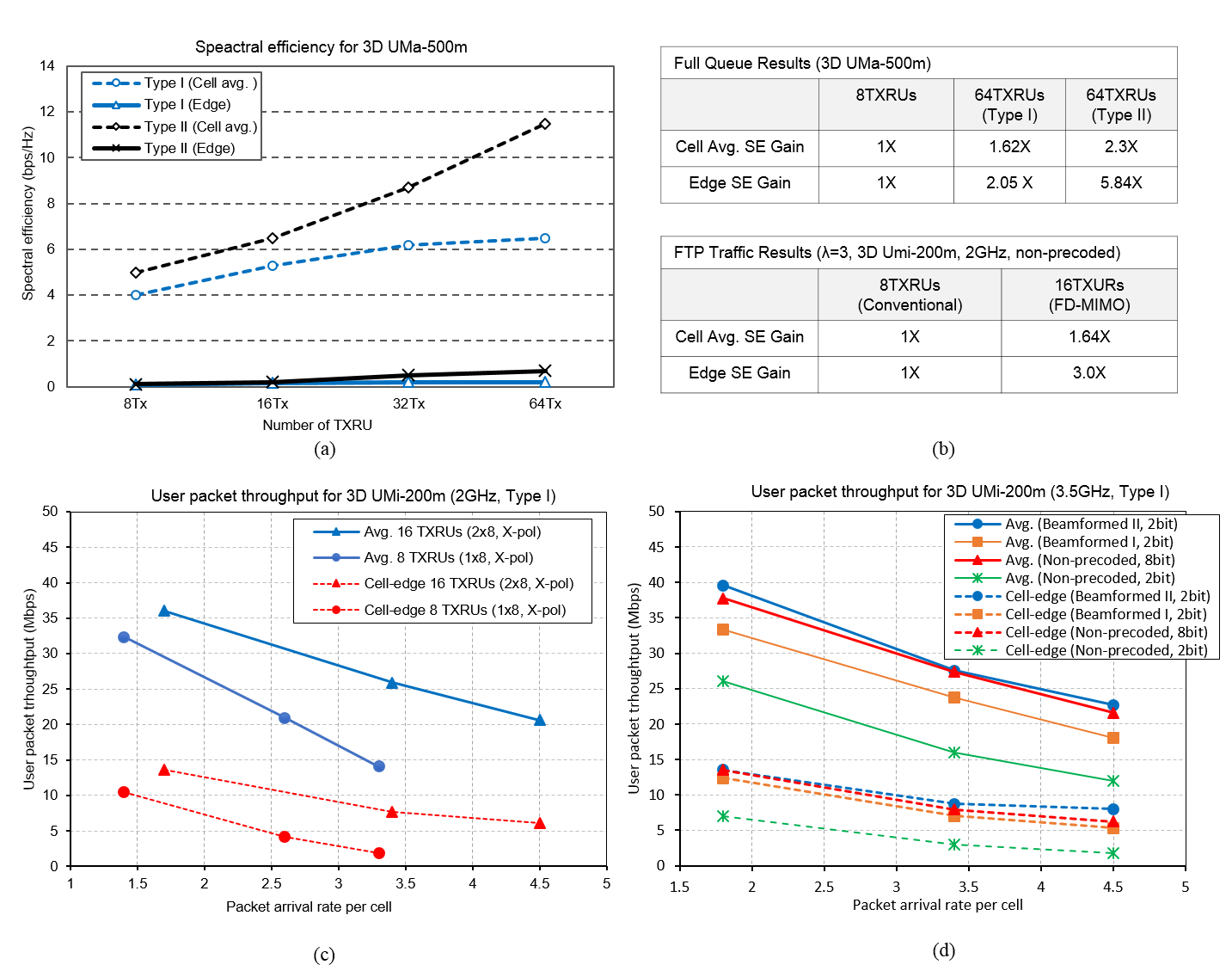}
		\caption{System level performance results and comparison with full buffer and FTP traffic model.}
		\label{fig:performance}
	\end{figure*}
	%	%%%%%%%%%%%%%%%%%%%%%%%%%%%%%%%%%%%%%%%%%%%%%%%%%%%%
	%Detailed simulation parameters are provided in Table \ref{tab:simulation}. 
%In Fig. \ref{fig:performance}(a),
%we plot the throughput of the conventional LTE systems with 8Tx ($N_V \times N_H = 1 \times 8$) and FD-MIMO systems with 16 and 32Tx ($N_V \times N_H = 2 \times 8, \ 4 \times 8$) for the full buffer traffic.
%
%Since the co-scheduling UEs in different floors of the same building results in lower interuser interference than the case of co-scheduling UEs in different buildings, FD-MIMO systems achieve substantial gain over the conventional LTE MIMO systems, resulting in $4$\% and $20$\% gain for cell average and $74$\% and $94$\% gain for cell edge of $16$Tx and $32$Tx, respectively.
%
%

In Fig. \ref{fig:performance}(c), we plot the user throughput of the finite traffic model as a function of packet arrival rate. Note that when the packet arrival rate is high, co-scheduled users need to be increased and thus the intercell and multiuser interference will also increase. In this realistic scenario, FD-MIMO systems outperform the conventional MIMO systems with a large margin, achieving 1.5${\mkern-2mu\times\mkern-2mu}$ and 3${\mkern-2mu\times\mkern-2mu}$ improvement in cell average and edge user packet throughput, respectively. Note that in the low network loading (low interference scenario), gain of the FD-MIMO systems is coming from the 3D beamforming. In the medium to high network loading (high interference scenario), this  gain is mainly due to the multiuser precoding of the 2D active antenna array. %\hl{This reveals that scheduling strategy of 2D AAS should consider the network traffic as well as UE locations and this study will be one of research directions.}  %This reveals that the 3D beamforming achieves SINR improvement for the low interference condition (low network loading) while MU precoding with 2D antenna array offers better spatial separation for the high interference condition (medium to high network loading).
Fig. \ref{fig:performance}(d) summarizes the throughput of various CSI feedback frameworks. With the same feedback overhead (2bit), beamformed CSI-RS scheme I outperforms the non-precoded scheme with a large margin. This is because the number of codewords for the channel feedback is only four so that channel state information at eNB is very coarse. %Since the number of precoder (number of codebook is 4) is not enough to obtain accurate channel, the results performs worse than the cases using beamformed CSI-RS. 
Since the beamformed CSI-RS scheme II can adapt weights of the beamformed CSI-RS to generate an accurate CDI, it performs best among all under tests. %better than other schemes. %ed from these and thus performance is better than other schemes. 
It is worth mentioning that the non-precoded CSI-RS scheme requires a large amount of feedback overhead (approximately 128 quantization levels) to achieve comparable performance to the CSI-RS scheme I. %,  and this will increase feedback overhead.
%Since the number of beamformed CSI-RSs ($N_B=4$) is much smaller than the number of the precoder (number of codebooks is 128) and further beamforming weights are fixed, it is not easy to obtain an accurate CDI for the beamformed transmission, resulting worst performance for both edge and cell average. Whereas, hybrid CSI-RS transmission can adapt weights of beamformed CSI-RS from the long-term CSI feedback, more accurate CDI can be obtained from beamformed CSI-RS and thus performance is better than other schemes even for the small $N_B$. %Hybrid transmission has benefits for reducing the overall CSI-RS resources and adapting the weight of beamformed CSI-RS.
%Those benefits can compensate performance degradation of beamformed transmission and provide even more gains in UEs.
{From this observation, we clearly see that the beamformed CSI-RS transmission is effective in controlling the precoding weights (in time, frequency, and space), feedback overhead, and pilot resource overhead. % and we expect that this scheme would be a key ingredient of the FD-MIMO systems. %will be indispensable paradigm in the future 5G massive antenna systems.} %Also, in combination of various TXRU architectures that we have yet to cover, there are open research areas of how to use beam in the wireless networks.}
%Nevertheless, as shown in Fig. \ref{fig:performance}(d), performance of beamformed transmission for the 5\%-tile user throughput is comparable to non-precoded, beamformed, and hybrid transmission.

%\begin{abstract}
%Multiple-input multiple-output (MIMO) systems with large number of basestation antennas, often called massive MIMO systems, have received much attention in academia and industry as a means to improve the spectral efficiency, energy efficiency, and also processing complexity.
%
%Mobile communication industry has initiated a feasibility study to meet the increasing demand of future wireless systems. Field trials for the proof-of-concept systems have demonstrated the potential gain of the FD-MIMO and 3rd generation partnership project (3GPP) standard body has initiated the standardization activity for the seamless integration of massive MIMO into current 4G cellular carrier frequency. A study item, process done before a formal standardization process, has been completed in June 2015 and the follow up (work item) process will be initiated shortly for the formal standardization of Release 13.
%
%In this article, we provides an overview of the FD-MIMO systems, with emphasis on the discussion and debate conducted on standardization process of Release 13. These include key features for FD-MIMO systems, summary of the major issues for the standardization and practical system design, performance evaluation for typical scenarios.
%\end{abstract}

	%%%%%%%%%%%%%%%%%%%%%%%%%%%%%%%%%%%%%%%%%%%%%%%%%%%%%%%%%%%%%%%%%%%%%%%%%%%
	\section{Concluding Remarks}
	%%%%%%%%%%%%%%%%%%%%%%%%%%%%%%%%%%%%%%%%%%%%%%%%%%%%%%%%%%%%%%%%%%%%%%%%%%%%
	\label{sec:conclusion}	
	In this article, we have provided an overview of FD-MIMO systems in 3GPP LTE (recently named as LTE-Advanced Pro) with emphasis on the discussion and debate conducted on the Rel. 13 phase. {We discussed key features of FD-MIMO systems and main issues in standardization of system design, such as channel model, transceiver architectures, pilot transmission, and CSI feedback scheme.} %We also presented the system level simulation results to demonstrate the potential gain of FD-MIMO systems.} % with various feedback scehems.
To make the most of a large number of eNB antennas in a cost and space effective manner, new key features, distinct from MIMO systems in conventional LTE-A, should be introduced in the standardization, system design, and transceiver implementation. These include new transmitter architecture (array connected architecture), new RS transmission scheme (beamformed CSI-RS transmissions), and enhanced channel feedback (beam index feedback). %, and many more.
%
 %success of commercial FD-MIMO systems and evolution to FDD based massive MIMO. We have still many technical issues remain unsolved and further study and investigation will be conducted through both the academia and the standardization process. % (work item phase).
Although our work focused primarily on the standardization in Rel. 13, there are still many issues for the successful deployment of FD-MIMO systems in the future, including pilot overhead reduction, beam adaptation and optimization, and advanced channel estimation exploiting time and angular domain sparsity \cite{bshim}. %, and RS overhead reduction. % for throughput enhancement.

\section*{Acknowledgments}
This work was supported by the National Research Foundation of Korea (NRF) grant funded by the Korean government (MSIP-2016R1A5A1011478) and the ICT R\&D program of MSIP/IITP (B0717-16-0023).

\clearpage
\begin{table}[t]
	\begin{center}
		\caption{System simulation assumptions}
		\label{tab:simulation}
		\begin{tabular}{l|l} \hline
			\textbf{Parameter} & \textbf{Value} \\ \hline
			Duplex method & FDD \\ \hline
			Bandwidth & 10 MHz \\ \hline
			Center frequency & 2GHz / 3.5GHz \\ \hline
			Inter-site distance & 500m for 3D-UMa, 200m for 3D-UMi \\ \hline
			Network synchronization & Synchronized \\ \hline
		    Cellular layout & 3D Hexagonal grid, 19 eNBs, 3 cells per site \\ \hline
			Users per cell & 10 (Uniformly located in 3D space) \\ \hline
			Downlink transmission scheme & $N_T \times 2$ MU-MIMO SLNR precoding with rank adaptation with 2 layer per UE\\ \hline
			Downlink scheduler & Proportional Fair scheduling in the frequency and time domain. \\ \hline
			Downlink link adaptation & CQI and PMI 5ms feedback period \\
			& 6ms delay total (measurement in subframe $n$ is used in subframe $n+6$) \\
			& Quantized CQI, PMI feedback error: 0$\%$ \\
			& MCSs based on LTE transport formats\\ \hline
			Downlink HARQ & Maximum 3 re-transmissions, IR, no error on ACK/NACK, 8ms delay between re-transmissions \\ \hline
			Downlink receiver type & MMSE : based on demodulation reference signal (DM-RS) of the serving cell \\ \hline
			Channel estimation & Non-ideal channel estimation on both CSI-RS and DM-RS\\ \hline
			Antenna configuration & $(M, N, P) = (8, 4, 2)$\\ \hline
			TXRU configuration ($N_H\times N_V$)& $1\times8$, $2\times8$, $4\times8$, and $8\times8$ with X-pol ($0.5\lambda$, $0.8 \lambda$ antenna spacing for vertical and horizontal)\\ \hline
			Control channel overhead, & Control channel: 3 symbols in a subframe\\
			Acknowledgments etc. & Overhead of DM-RS: 12 RE/RB/Subframe \\
			& Overhead of CSI-RS: in maximum 16 REs of CSI-RS every 5ms per RB \\
			& (This is, in 8 Tx antenna case, 8 REs/RB per 10ms)\\
			& Overhead of CRS: 2-ports CRS \\ \hline
			Channel model & 3D urban macro and micro channel model \cite{3D} with 3km/h UE speed \\ \hline
			Inter-cell interference modeling & 57 intercell interference links are explicitly considered. \\ \hline
			Max. number of layers & $4$ \\ \hline
			Traffic model & Full buffer and non-full buffer (FTP Model) with 0.5 MBytes packet and various arrival rate \\ \hline
		\end{tabular}
	\end{center}
\end{table}

\begin{biography}{Hyoungju Ji}
	is currently working toward the Ph.D. degree  School of Electrical and Computer Engineering, Seoul National University, Seoul, Korea. He joined Samsung Electronics in 2007, and has been involved in 3GPP RAN1 LTE technology developments and standardization.
	His current interests include multi-antenna techniques, massive connectivity, machine type communications, and IoT communications.
\end{biography}

\vspace{-0.7cm}
\begin{biography}{Younsun Kim}
	received B.S. and M.S. degrees in electronic engineering from Yonsei University, Korea, and his Ph.D. degree in electrical engineering from the University of Washington, in 1996, 1999, and 2009, respectively. He joined Samsung Electronics in 1999
	and has been working on the standardization of wireless communication systems such as cdma2000, HRPD, and recently LTE/LTE-A. His research interests include multiple access schemes, coordination schemes, multiple antenna techniques, and advanced receivers for next generation systems. 
\end{biography}

\vspace{-0.7cm}
\begin{biography}{Juho Lee}
	is currently a Master (technical VP) with Samsung Electronics and is in charge of research on standardization of wireless communications. He received his B.S., M.S., and Ph.D. degrees in electrical engineering from Korea Advanced Institute of Science and Technology (KAIST), Korea, in 1993, 1995, and 2000, respectively. He joined Samsung Electronics in 2000 and has been working on standardization of mobile communications for 3G and 4G such as WCDMA, HSDPA, HUSPA, LTE, and LTE-Advanced and is also actively working on research and standardization for 5G. He was a vice chairman of TSG RAN WG1 during February 2003 through August 2009, chaired LTE/LTE-Advanced MIMO sessions, and served as the rapporteur for the 3GPP LTE-Advanced Rel-11 CoMP work item.
\end{biography}

\vspace{-0.7cm}
\begin{biography}{Eko Onggosanusi}
	is currently a director of standards at the Standardization and Multimedia Innovation (SMI) Lab of Samsung Dallas. Prior to joining Samsung in 2014, he was a manager at Texas Instruments, working on cellular standards and algorithm/system designs especially HSPA and LTE systems. Having been a 3GPP RAN1 delegate since 2005, he has contributed to numerous components of LTE physical layer specification. He was the 3GPP rapporteur of the EBF/FD-MIMO study and work items and is currently the 3GPP rapporteur of the Enhanced FD-MIMO work item. He received his Ph.D. in electrical engineering from the University of Wisconsin-Madison (2000), has authored a number of papers in conferences and peer-reviewed journals, and is an inventor of numerous patents in wireless communications.
\end{biography}

\vspace{-0.7cm}
\begin{biography}{Younghan Nam}
	is currently a Senior Staff Engineer in Samsung Research America, Richardson TX. He has been engaged in standardization, design and analysis of the 3GPP LTE, LTE-Advanced and 5G NR since 2008. He is currently a study item rapporteur of the above 6GHz channel models (3GPP TR38.900). He received a Ph.D. in electrical engineering from the Ohio State University, Columbus OH, in 2008, and received his M.S. and B.S from Seoul National University, Korea, in 2002 and 1998, respectively. His research interests include MIMO, cooperative communications, and channel modeling.
\end{biography}

\vspace{-0.7cm}
\begin{biography}{Jianzhong Zhang}
	is a VP and head of Standards and Mobility Innovation Lab with Samsung Research America, where he leads research and standards for 5G cellular systems and next generation multimedia networks. He received his Ph.D. degree from University of Wisconsin, Madison.  From August 2009 to August 2013, he served as the Vice Chairman of the 3GPP RAN1 working group and led development of LTE and LTE-Advanced technologies such as 3D channel modeling, UL-MIMO and CoMP, Carrier Aggregation for TD-LTE, etc. Before joining Samsung, he was with Motorola from 2006 to 2007 working on 3GPP HSPA standards, and with Nokia Research Center from 2001 to 2006 working on IEEE 802.16e (WiMAX) standard and EDGE/CDMA receiver algorithms. Dr. Zhang is a Fellow of IEEE.”
\end{biography}

\vspace{-0.7cm}
\begin{biography}{Byungju Lee}
	received the B.S. and Ph.D. degrees in the School of Information and Communication, Korea University, Seoul, Korea, in 2008 and 2014, respectively. He is now with the School of Electrical and Computer Engineering at Purdue University, West Lafayette, IN, USA, as a postdoctoral scholar. From 2014 to 2015, he was a postdoctoral fellow at Seoul National University, Seoul, Korea. %He worked as a visiting researcher with Purdue University, West Lafayette, IN, USA, in 2013. 
	His research interests include information theory and signal processing for wireless communications.
\end{biography}

\vspace{-0.7cm}
\begin{biography}{Byonghyo Shim}
	is an associate professor in the Electrical and Computer Engineering at the Seoul National University, and a director of information system laboratory. He received B.S. and M.S. degrees in control and instrumentation engineering from Seoul National University in 1995 and 1997, respectively, and M.S. degree in mathematics and Ph.D. degree in electrical and computer engineering from the University of Illinois at Urbana-Champaign in 2004 and 2005, respectively. From 2005 to 2007, he worked for Qualcomm Incorporated and from 2007 to 2014, he was with Korea University. His current research focuses on 5G wireless communications (physical layer system design) and bigdata signal processing.
\end{biography}

\end{document}